\documentclass[final,1p,times]{elsarticle}

\usepackage{graphicx}
\usepackage{amssymb}
\usepackage[T1]{fontenc}
\journal{Nuclear Physics A}

\begin{document}

\begin{frontmatter}

\title{Strangeness in Compact Stars\tnoteref{hyp2009}}

\tnotetext[hyp2009]{invited talk given at the 10th International Conference on
    Hypernuclear and Strange Particle Physics (HYP-X) September 14-18,
    2009, Tokai, Ibaraki, Japan}

\author{J\"urgen Schaffner-Bielich}

\address{Philosophenweg 16,\\ Institute for Theoretical Physics,\\
  University of Heidelberg,\\ D-69120 Heidelberg, Germany}

\ead{schaffner-bielich@uni-heidelberg.de}

\begin{abstract}
  We discuss the impact of strange hadrons, in particular hyperons, on
  the gross features of compact stars and on core-collapse supernovae.
  Hyperons are likely to be the first exotic species which appears
  around twice normal nuclear matter density in the core of neutron
  stars. Their presence largely influences the mass-radius relation of
  compact stars, the maximum mass, the cooling of neutron stars, the
  stability with regard to the emission of gravitational waves from
  rotation-powered neutron stars and the possible early onset of the
  QCD phase transition in core-collapse supernovae. We outline also
  the constraints from subthreshold kaon production in heavy-ion
  collisions for the maximum possible
  mass of neutron stars.
\end{abstract}


\end{frontmatter}

%%%%%%%%%%%%%%%%%%%%%%%%%%%%%%%%%%%%%%%%%%%%%%%%%%%%%%%%%%%%%%%%%%%%%%%%%%%

\section{Introduction}

%%%%%%%%%%%%%%%%%%%%%%%%%%%%%%%%%%%%%%%%%%%%%%%%%%%%%%%%%%%%%%%%%%%%%%%%%%%

Neutron stars constitute a fantastic laboratory for studying matter
under extreme conditions. In particular in the core of neutron stars,
new exotic phases could be present with considerable impacts on its
evolution and global properties. The focus of the following discussion
will be on the presence of hyperons being a major component
of the composition of neutron star matter at high densities which is
largely based on the recent review \cite{SchaffnerBielich:2008kb}. 

The maximum masses of neutron stars are controlled by the stiffness of
the nuclear equation of state which is related to the 
three-body force involving hyperons. The subthreshold
production of kaons in heavy-ion collisions provide a new limit on the
maximum possible mass of neutron stars which just relies on the
nuclear equation of state, as constraint by the heavy-ion data, up to
a fiducial density and causality arguments. The cooling of neutron
stars for up to about a million years is governed by the emission of
neutrinos. Here the weak processes involving hyperons allow for fast
cooling depending on the size of the hyperon gap, i.e.\ the two-body
interaction between hyperons. Rotating neutron stars can emit
gravitational waves due to the presence of the so called r-mode
instability which is highly sensitive to the viscosity of dense
matter. Nonmesonic weak processes with hyperons turn out to be a
decisive ingredient for the viscosity and therefore for the stability
of rotation-powered neutron stars, in particular for hot proto-neutron
stars or accreting binary rotation-powered neutron stars. As hyperons
appear at moderate densities, about twice normal nuclear matter
density for cold neutron star matter, they are present to some amount
also in hot supernova matter shortly after the bounce. Fluctuations
can trigger then more easily the nucleation process for forming
bubbles of strange quark matter thereby enabling the onset of the
QCD phase transition much earlier in the evolution of the supernova,
maybe already shortly after the bounce. A first order phase transition
present during the first second of a core-collapse supernova has
profound implications for the overall evolution of the supernova, the
neutrino signal and possibly for the r-process nucleosynthesis in the
neutrino wind of the proto-neutron star.

%%%%%%%%%%%%%%%%%%%%%%%%%%%%%%%%%%%%%%%%%%%%%%%%%%%%%%%%%%%%%%%%%%%%%%%%%%%

\section{Observations of neutron stars}

%%%%%%%%%%%%%%%%%%%%%%%%%%%%%%%%%%%%%%%%%%%%%%%%%%%%%%%%%%%%%%%%%%%%%%%%%%%

Neutron Stars are extremely compact, massive objects with typical
radii of $\approx$ 10 km and typical masses of $1-2M_\odot$. Hence,
extreme densities have to be present in the core of a neutron star,
which must be several times nuclear density $n\gg n_0 = 3\cdot
10^{14}$ g/cm$^3$.  Masses of pulsars can be determined to quite some
precision by the observation of binary pulsars, i.e.\ rotation-powered
neutron stars with a companion which is either a normal star, a white
dwarf or even another neutron star.

The best determined mass of$M=(1.4414\pm 0.0002)M_\odot$ for the
Hulse-Taylor pulsar \cite{Weisberg:2004hi} still constitutes the most
massive known neutron star mass which is well known and relies just on
post-Keplerian analysis of the orbital parameters (note that the
mass of PSR J0751+1807 has been corrected from $M=(2.1\pm 0.2)
M_\odot$ to $M=(1.14-1.40)M_\odot$ \cite{Nice:2008}).  A new
measurement of the pulsar PSR J1903+0327 arrives at a mass of $M=
(1.67\pm 0.01) M_\odot$ \cite{Champion:2008,Freire:2009fn} but the
improved mass measurement is not finalized yet.

Constraints on the mass-radius relation (see e.g.\
\cite{Lattimer:2006xb} for a review) can be derived from the spin rate
from PSR B1937+21 of 641 Hz which gives a radius of $R<15.5$ km for
$M=1.4M_\odot$ The causality limit for the nuclear equation of state
with $R>3GM$ bounds the range for the possible mass-radius relation
from the other side, so that a distorted pie-like region remains. We
stress that any further constraint on the mass-radius relation of
neutron stars discussed so far in the literature is liable to suffer
from particular model assumptions, so we refrain from discussing them
here and refer to the above mentioned review.

%%%%%%%%%%%%%%%%%%%%%%%%%%%%%%%%%%%%%%%%%%%%%%%%%%%%%%%%%%%%%%%%%%%%%%%%%%

\section{Composition of Neutron Stars}

%%%%%%%%%%%%%%%%%%%%%%%%%%%%%%%%%%%%%%%%%%%%%%%%%%%%%%%%%%%%%%%%%%%%%%%%%%

The structure of a neutron star in the core is not known at
present. Several scenarios have been discussed in the literature as
e.g.\ the formation of a pion condensed phase, kaon condensation, the
transition to strange quark matter and the possibility of having pure
strange quark stars (for a review see \cite{Weber:2004kj}). We will
argue in the following that the first exotic phase which appears in
the core of neutron stars is likely to be hyperonic in nature.

\begin{table}
\begin{center}
\begin{tabular}{c|c|c|c|c}
Hadron        & p,n & $\Sigma^-$ & $\Lambda$ & others \cr
\hline
appears at: & $\ll n_0$ & $4n_0$ & $8n_0$ & $>20 n_0$
\end{tabular}
\end{center}
\caption{The critical density for the appearance of hadrons in neutron
star matter for a free gas. Hyperons are present at $4n_0$ but no pion or
kaon condensation is formed for densities below $20n_0$.}
\label{table:onset}
\end{table}

The general critical condition for the presence of any particle in
equilibrated cold matter is that the chemical potential of the
particle equals its in-medium energy (note that this assumes that one
can adopt a quasi-particle picture).  Neutron star matter is in
$\beta$-equilibrium, so that all weak decays are Pauli-blocked. Hence,
neutrons and hyperons, if present, can not decay by weak interactions
as in free space.  For boson condensation (in the s-wave) of $\pi^-$
and $K^-$ the critical condition reads $E_b = \mu_b= \mu_e$.  For a
free gas of electrons, muons and hadrons one finds in
table~\ref{table:onset} that the $\Sigma^-$ is the first strange
hadron which is present at high densities appearing at $4n_0$,
followed by the $\Lambda$ at $8n_0$ \cite{Ambart60}. The $\Sigma^-$ is
favoured due to its negative charge which balances the positive charge
of the protons and helps to reduce the Fermi energy of the
electrons. Note that all other hadrons, as the $\Sigma^+$, the
$\Xi^-$, the $\pi^-$ or the $K^-$ are not part of the composition for
any reasonable nuclear density, $n<20n_0$.  Note that interactions
will considerably change the critical density for the onset of a
species.  But the corresponding equation of state results in a maximum
mass of only $ M_{\rm max} \approx 0.7 M_\odot < 1.44 M_\odot $
\cite{OV39} which implies that effects from strong interactions are
essential to describe pulsar data and therefore the gross features of
neutron stars.

There have been numerous models being utilized to determine the onset
of hyperons in neutron star matter. If these models are properly
adjusted to the available hypernuclear data, they find 
consistently that hyperons appear around $n\approx 2n_0$.
Such investigations include 
relativistic mean-field models
\cite{Glendenning:1984jr,Ellis:1995kz,Knorren:1995ds,Schaffner:1995th}, 
the nonrelativistic potential model \cite{Balberg:1997yw},
the quark-meson coupling model \cite{Pal:1999sq},
the relativistic Hartree--Fock approach \cite{Huber:1998hm},
Brueckner--Hartree--Fock calculations
\cite{Baldo:1998hd,Baldo:1999rq,Vidana:2000ew,Schulze:2006vw}, 
chiral effective Lagrangians using SU(3) symmetry
\cite{Hanauske:1999ga,Schramm:2002pa,Dexheimer:2008ax},
the density-dependent hadron field theory \cite{Hofmann:2000mc},
G-matrix calculations \cite{Nishizaki:2002ih} and
the renormalization group approach with a $V_{\rm low~k}$ potential
for hyperons \cite{Djapo:2008au}. Hence, most likely 
neutron stars are giant hypernuclei \cite{Glendenning:1984jr}!

The composition of neutron star matter is largely controlled by the
hyperon potential depths which are fixed to the available hypernuclear
data and hadronic atom data. For an attractive potential of the
$\Sigma$ hyperons, the $\Sigma^-$ can appear even before the $\Lambda$
in dense matter. But if the $\Sigma$-nucleon potential is repulsive at
normal nuclear matter density, then the $\Sigma$ hyperons are not
populated at all, only the $\Lambda$s are present in matter around
$n=2n_0$, the $\Xi^-$ before $n=3n_0$.  Therefore, the hyperon
population is highly sensitive to the in-medium potential.

%%%%%%%%%%%%%%%%%%%%%%%%%%%%%%%%%%%%%%%%%%%%%%%%%%%%%%%%%%%%%%%%%%%%%%%%%%%

\section{Masses of Neutron Stars}

%%%%%%%%%%%%%%%%%%%%%%%%%%%%%%%%%%%%%%%%%%%%%%%%%%%%%%%%%%%%%%%%%%%%%%%%%%%

Hyperons can also play a crucial role for the maximum mass of neutron
stars, which is controlled by the high-density equation of state. The
more new particles emerge at high densities, the more can the Fermi energy
be lowered for a given number density which results in a lowered Fermi
pressure for a given energy density for nonrelativistic
particles. Note that this argument can not be applied to completely
new phases as e.g.\ the chirally restored phase (quark matter).

\begin{figure}
\centerline{\includegraphics[width=0.5\textwidth]{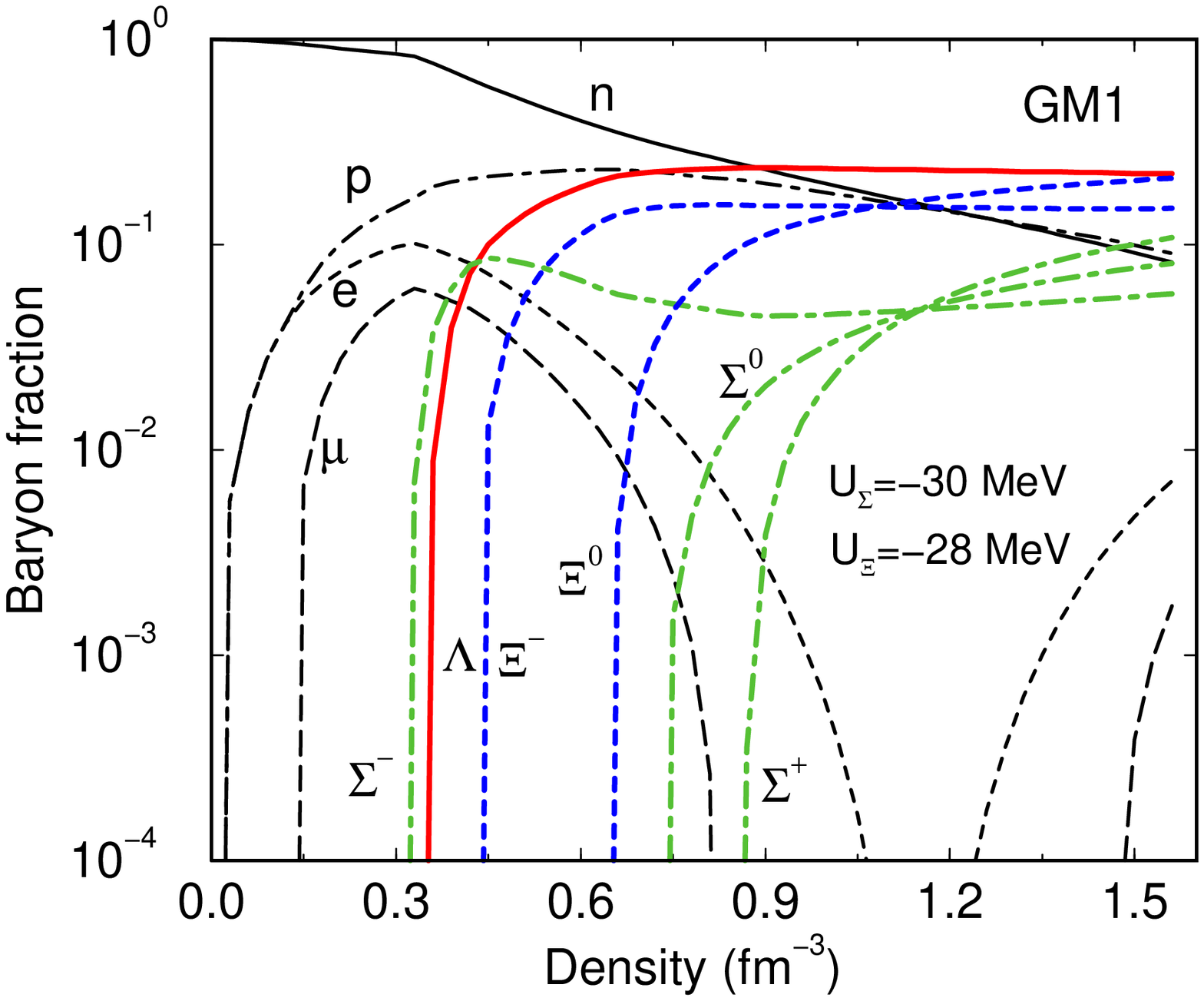}
\includegraphics[width=0.5\textwidth]{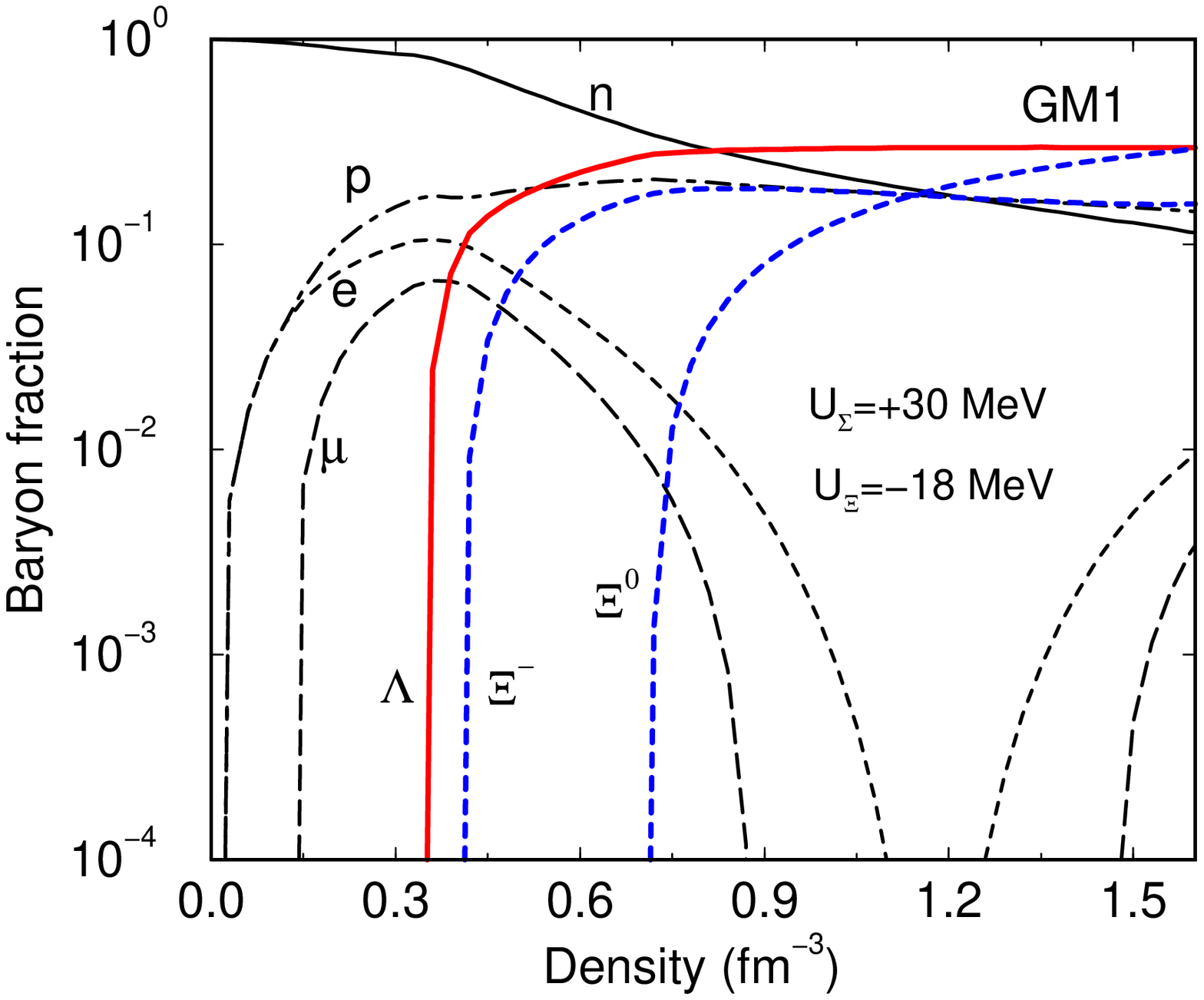}}
\caption{The composition of neutron star matter as a function of baryon
  density. Hyperons appear around $2n_0$. The presence of the $\Sigma$
  hyperons depends crucially on the sign of the hyperon-nucleon
  potential, there are no $\Sigma$ hyperons present for a repulsive
  potential. Left plot: attractive $\Sigma$ potential, right plot:
  repulsive $\Sigma$ potential (see \cite{Schaffner:1995th} for the
  details of the model used).}
\end{figure}

The first discussion of the implication for the maximum mass of
neutron stars with hyperons which takes into account correctly the
hyperon potential depths in dense matter is given in
\cite{Glendenning:1991es} within the relativistic mean-field
model. While the maximum mass of neutron stars with nucleons and
leptons only is about $M\approx 2.3M_\odot$, it drops substantial due
to hyperons.  The maximum mass for ``giant hypernuclei'' reads only
$M\approx 1.7 M_\odot$, noninteracting hyperons result in a too low
mass of $M<1.4M_\odot$ which is incompatible with observations. Hence,
at high densities repulsive interactions between hyperons and nucleons
are important for the stability of neutron stars.

Modern many-body approaches use as input the two-body potentials as 
deduced from hyperon-nucleon scattering data.  For the Nijmegen
soft-core hyperon nucleon potentials Vidana et al. find that the
maximum mass is only $M_{\rm max} = 1.47 M_\odot$ which reduces to
even $M_{\rm max} = 1.34 M_\odot$ when the hyperon-hyperon potentials
are switched on \cite{Vidana:2000ew}. Also Baldo et al. find values of
$M_{\rm max} = 1.26 M_\odot$ even when including three-body nucleon
interactions \cite{Baldo:1999rq}. More recently Schulze et al.
\cite{Schulze:2006vw} and Djapo et al. \cite{Djapo:2008au} confirm
that $M_{\rm max} < 1.4 M_\odot$ for modern microscopic (ab initio)
approaches. Hence, the neutron star equation of state gets too soft at
high densities giving too low masses. Probably the underlying reason
are missing three-body forces for hyperons (YNN, YYN, YYY), which give
additional repulsive contributions at high densities.  If so then it
seems that neutron stars can not live without hyperon three-body
forces. A solution to this problem of the modern many-body approaches
has been discussed in detail in ref.~\cite{Baldo:2003vx} where it was
shown that the presence of quark matter in the core stabilizes the
compact star resulting in much larger maximum masses.  But certainly,
also more input is needed from hypernuclear physics by e.g.\ the study
of light double hypernuclei in the near future to extract the
hyperonic three-body forces.

%%%%%%%%%%%%%%%%%%%%%%%%%%%%%%%%%%%%%%%%%%%%%%%%%%%%%%%%%%%%%%%%%%%%%%%%%%%

\section{Maximum possible mass of neutron stars}

%%%%%%%%%%%%%%%%%%%%%%%%%%%%%%%%%%%%%%%%%%%%%%%%%%%%%%%%%%%%%%%%%%%%%%%%%%%

There is another strange hadron with strong relations to the physics
of the maximum possible mass of neutron stars.  Kaons produced
subthreshold in heavy-ion experiments can serve as a messenger of the
high-density zone created in the collision.  Kaons are produced by
associated production e.g.\ via NN$\to{\rm N}\Lambda$K, and
NN$\to$NNK$\overline{\rm K}$ in elementary proton-proton
collisions. In the medium, i.e.\ in heavy-ion collisions, rescattering
processes open up as $\pi{\rm N}\to\Lambda$K, $\pi\Lambda\to{\rm
  N}\overline{\rm K}$ from produced pions which have a lower q-value
and are therefore able to pump up the kaon production rates
substantially compared to the elementary pp-collisions.  At
subthreshold bombarding energies of heavy ions the matter can be
compressed up to $3n_0$. However, kaons have a long mean-free path,
they scatter elastically with nucleons and pions, only hyperons can
absorb them as kaons carry an antistrange quark. Hence, kaons can
escape from the high density zone unimpeded, their production rates
will be controlled by the amount of compression achieved in the
central region of the collision. For a recent review of the relation
between the nuclear equation of state and kaon production rates in
heavy-ion collisions see \cite{Aichelin:2008}.

The KaoS collaboration has measured kaon production ($K^+$) in
heavy-ion collisions at subthreshold energies
\cite{Sturm:2000dm,Forster:2007qk}. They used carbon-carbon collision
as a control experiment to assess the medium effects in comparison to
heavy-ion collisions with gold-gold collisions at 0.8 AGeV and 1.0
AGeV. The multiplicity per mass number for Au+Au collisions relative
to the one for C+C collisions turns out to be rather insensitive to
input parameters in numerical simulations as effects from two-body
interactions, cross sections, and in-medium potentials effectively
cancel out \cite{Fuchs:2000kp,Hartnack:2005tr}. A strong increase of
the kaon production rate was seen towards lower collision energies
which could only be matched by transport calculations with a soft
nuclear equation of state, here characterized by a compression modulus
of $K_N\approx 200$ within a simple Skyrme parametrization.

These findings can now be utilized for constraining the maximum
possible mass of neutron stars using causality arguments.  Let us
assume that we know the nuclear equation of state up to some fiducial
density $\epsilon_f$. Then the neutron star matter can not be stiffer
than the causality limit which is $p=\epsilon$ above that fiducial
density as shown by Rhoades and Ruffini \cite{Rhoades:1974fn}. Hence,
there is a maximum mass possible which is related to the fiducial
density by $M_{\rm max}= 4.2M_\odot (\epsilon_0/\epsilon_f)^{1/2}$,
where $\epsilon_0$ is the energy density of nuclear matter at
saturation (see e.g.\ \cite{Kalogera:1996} who are using a nuclear
equation of state derived from fits to nuclei and low-energy
nucleon-nucleon scattering data). As the new constraint on the nuclear
equation of state as determined from the analysis of the KaoS data
concerns densities above normal nuclear matter density, the limit on
the maximum mass can be lowered accordingly by increasing the fiducial
density to $\epsilon_f\approx 2\epsilon_0$
\cite{Sagert:2007nt,Sagert:2007kx} arriving at a new upper mass limit
of about $2.7 M_\odot$ from heavy-ion data.

%%%%%%%%%%%%%%%%%%%%%%%%%%%%%%%%%%%%%%%%%%%%%%%%%%%%%%%%%%%%%%%%%%%%%%%%%%%%

\section{Weak Hyperonic Reactions and Neutron Stars}

%%%%%%%%%%%%%%%%%%%%%%%%%%%%%%%%%%%%%%%%%%%%%%%%%%%%%%%%%%%%%%%%%%%%%%%%%%%%%

A neutron stars cools most efficiently by emitting neutrinos for the
first million years after being created in a core-collapse supernova.
The modified URCA process occurs at finite density and is slow as it
needs a bystander nucleon to conserve energy and momentum: $N + p +
e^- \to N + n + \nu_e $ and $N + n \to N + p + e^- + \bar\nu_e$.
Without a bystander nucleon, the so called direct URCA process is much
faster due to the increased phase space: $ p + e^- \to n + \nu_e$ and
$n \to p + e^- + \bar\nu_e $ but can only proceed for $p_F^p + p_F^e
\geq p_F^n$ to conserve energy and momentum. As charge neutrality
implies that $n_p = n_e$ the proton fraction must be at least $n_p/n
\geq 1/9$ so that the nucleon URCA process is only allowed for large
proton fractions.  On the contrary the hyperon URCA process as $\Lambda
\to p + e^- + \bar\nu_e$ and $\Sigma^- \to n + e^- + \bar\nu_e$ 
happen immediately when the respective hyperons are present. These
reactions enable fast cooling and are only suppressed by hyperon
pairing gaps. Cooling with hyperons has been studied in refs.\ 
\cite{Schaab:1998zq,Page:2000wt,Vidana:2004rd,Zhou:2005hv,Takatsuka:2005bp}
where two-body hyperon-hyperon interactions were used as input for
the calculations of the neutron star cooling rates. Pairing of
$\Sigma$ hyperons for cooling processes in neutron stars were studied
by Vidana and Tolos in \cite{Vidana:2004rd}. Generically, the cooling
depends crucially on the composition of the neutron star, in
particular whether or not hyperons are present which will be the case
beyond a certain critical neutron star mass. To really assess the role
of hyperons for the cooling mechanism of neutron stars one needs to
know also the neutron star mass. Unfortunately, up to now the masses
have not been determined for those nearby and young neutron stars
where the luminosity in x-rays has been measured and constraints on
the cooling curves could be extracted.

There is another effect which is dominantly controlled by weak
reactions involving hyperons: the gravitational wave emission from
rotating neutron stars by the so called r-mode instability
\cite{Andersson:1997xt,Jones:2001ie,Jones:2001ya,Lindblom:2001hd,Haensel:2001em,vanDalen:2003uy,Drago:2003wg,Nayyar:2005th,Chatterjee:2006hy,Gusakov:2008hv}.
Oscillations of the neutron star brings the matter out of
$\beta$-equilibrium, as there are overdense and underdense regions.
The dominating effect to restore equilibrium is by weak nonmesonic
processes of the kind $NN\leftrightarrow \Lambda N$ and
$NN\leftrightarrow \Sigma N$. Strong reactions are faster but they can
not change strangeness to reestablish weak equilibrium in the neutron
star material. If those weak reactions are not taken into account, the neutron star can
not damp the oscillations, has to emit gravitational waves and slows
down. This feature creates an instability window for certain
combinations of the temperature and rotation frequency of the star. The key
ingredient for the stability relative to the emission of gravitational
waves is the viscosity which depends crucially on hyperon
weak nonmesonic reactions. If hyperons are gapped these reactions are
suppressed, so that the hyperon-hyperon interactions play again an
important role, see also \cite{Chatterjee:2006hy}. Recently, the LIGO
collaboration has published new limits on the gravitational wave
emission from the Crab pulsar which are well below the spin-down limit
and constrain already the amount of energy which can be emitted by
gravitational waves substantially \cite{Collaboration:2009rfa}.

%%%%%%%%%%%%%%%%%%%%%%%%%%%%%%%%%%%%%%%%%%%%%%%%%%%%%%%%%%%%%%%%%%%%%%%%%%%

\section{Hyperons and the QCD phase transition in supernovae}

%%%%%%%%%%%%%%%%%%%%%%%%%%%%%%%%%%%%%%%%%%%%%%%%%%%%%%%%%%%%%%%%%%%%%%%%%%%%

Finally, we address the importance of hyperon populations for the QCD
phase transition in core-collapse supernova explosions.  Stars with a
mass of more than eight solar masses end in a core-collapse
supernova. In recent years new generations of simulation codes have
been developed which includes multidimensional treatments and improved
approximations for the neutrino transport. Still, the shock front
stalls and can only be reinvigorated by neutrino heating for low
progenitor masses. The underlying mechanism for massive progenitor
stars to explode has not yet fully agreed upon and several mechanism
have been proposed as e.g.\ the standing accretion shock instability
(for a review see \cite{Janka:2006fh}). 

Hyperons can also be present in supernova matter as densities above
saturation densities and temperatures of about 20 MeV are achieved
shortly after the bounce.  A supernova matter equation of state with
hyperons has been studied by Ishizuka et al.~\cite{Ishizuka:2008gr}
recently. For a proton fraction of $Y_p=0.4$ and a temperature of 20
MeV the population of hyperons reaches about 0.1\%. Net strangeness is
produced thermally as hyperons are in weak equilibrium. The presence
of hyperons softens the nuclear equation of state, so that the
recollapse of massive progenitor stars of 100$M_\odot$ to a black
hole is triggered \cite{Sumiyoshi:2008kw}. 

Fluctuations in the hyperon abundances help to form local regions with
a high strangeness content. Then, it is much more feasible to nucleate
those regions to strange quark matter directly than to normal quark
matter as demonstrated by Mintz et al.~\cite{Mintz:2009ay}. Therefore,
the presence of hyperons catalyzes substantially the production of
strange quark matter bubbles allowing for the onset of the QCD phase
transition in supernova matter.  The conditions in supernovae are
favourable for the QCD phase transition to occur: quark matter appears
at much lower density due to weak equilibrium, the low critical
density for low proton fractions due to the nuclear symmetry energy
and the finite temperature.  For the situation in heavy-ion
collisions, the phase transition line at low temperatures and high
baryochemical potentials is located at much higher densities as there
is no weak equilibrium so that normal quark matter has to be produced
initially and due to the isospin-symmetric matter present.  Hence, for
the supernova matter at bounce with $T=10-20$ MeV, $Y_p=0.2-0.3$,
$\epsilon=(1-1.5)\epsilon_0$ production of quark matter in
supernovae at bounce seems to be feasible \cite{Sagert:2009bn} without
any contradiction to heavy-ion data.  The implications of an early
onset of the QCD phase transition for core-collapse supernovae are that
a second shock wave is produced which releases a second burst of
antineutrinos when the shock front is running over the neutrinosphere
\cite{Sagert:2008ka}. The neutrino signal of the phase transition
shows up in the temporal profile of the emitted neutrinos from the
supernova.  There is a pronounced second peak of anti-neutrinos due to
the formation of quark matter whose peak location and height is
determined by the critical density and strength of the QCD phase
transition \cite{Sagert:2008ka}.

%%%%%%%%%%%%%%%%%%%%%%%%%%%%%%%%%%%%%%%%%%%%%%%%%%%%%%%%%%%%%%%%%%%%%%%%%%%%

\section{Summary}

%%%%%%%%%%%%%%%%%%%%%%%%%%%%%%%%%%%%%%%%%%%%%%%%%%%%%%%%%%%%%%%%%%%%%%%%%%%%

Hypernuclear physics has a substantial impact on neutron star
properties.  Two-body hyperon-nucleon interaction controls the
composition of neutron star matter. Hyperons are most likely the first
exotic phase to appear in the core around twice normal nuclear matter
density. Hyperons can pair and form superfluids or superconducting
phases. The three-body hyperon-nucleon and hyperon-hyperon forces are
important for the maximum mass of neutron stars. Only low maximum
masses below $1.4 M_\odot$ are found in modern approaches without the
hyperonic three-body force. Kaon production in heavy-ion collision are
a probe of the nuclear equation of state at subthreshold energies. The
experimental data sets a new upper limit on the maximum mass allowed
by causality.  Nonmesonic weak reactions with hyperons are crucial for
the cooling history of young neutron stars as hyperons can cool
neutron stars rapidly by the direct hyperon URCA process. Also, weak
reactions with hyperons damp the r-mode instability of rotating
neutron stars and their gravitational wave emission. In those latter
two cases, hyperon pairing will affect those cooling rates and the
viscosity of dense neutron star matter. Finally, hyperons can be
produced thermally in supernova matter so that there is a finite
amount of strangeness present which can trigger the phase transition
to quark matter. A first order QCD phase transition can be read off
from the neutrino spectrum by a pronounced second peak in
antineutrinos emitted from a galactic supernova.

%%%%%%%%%%%%%%%%%%%%%%%%%%%%%%%%%%%%%%%%%%%%%%%%%%%%%%%%%%%%%%%%%%%%%%%%%%%%

This work is supported by the German Research
Foundation (DFG) within the framework of the excellence initiative
through the Heidelberg Graduate School of Fundamental Physics.
I thank David Blaschke for bringing the work of ref.~\cite{Baldo:2003vx}
to my attention.

\bibliographystyle{utphys}
\bibliography{all,literat}

\end{document}